# Voltage Controlled Resistance

Ashima Katyal, Parul Gupta and P. Arun*
*Department of Physics and Electronics*
*S.G.T.B. Khalsa College, University of Delhi, Delhi 110007*
*arunp92@physics.du.ac.in

**ABSTRACT**

*The paper discusses an application of the field effected transistor (FET) as a voltage controlled resistance which can be done in under-graduate labs either as a routine experiment or as a project.*

The *I-V* characteristics of a field effect transistor (FET) is often done as a routine experiment in the undergraduate physics and electronics laboratories. In theory class the advantages of this device over the Bipolar junction transistor (BJT) is also recalled. An important aspect of FETs is that their input impedance is far greater than that of BJTs, with the result they draw less current and in turn consume less power. Thus, FET are ideally suited for products that run on battery or cells. Also, while the load current (the collector current) in a BJT is controlled by the base current, FETs are voltage controlled devices. That is, the load current (Drain-Source current) of a FET is controlled by the voltage applied on the gate pin.

An important aspect that students miss (since it is not explicitly stated) is that, on changing the voltage applied on the FET's gate pin the resistance offered by the FET's channel changes. Different values of resistance draw different currents from the circuit and hence the statement "*FET are voltage controlled devices*". The channel of an FET is a semiconductor of length *l* and cross sectional area $w^2$, where we assume the FET to have a square cross-section of width *w* (Figure 1). On reverse biasing, the gate with respect to the source, a depletion width appears between the p- and n- semiconductors. This effectively decreases the cross-sectional area available for conduction and hence, the channel's resistance increases. In other words:

$$R_{\text{channel}} = \rho \left[ \frac{l}{w^2(V_{GS})} \right]$$

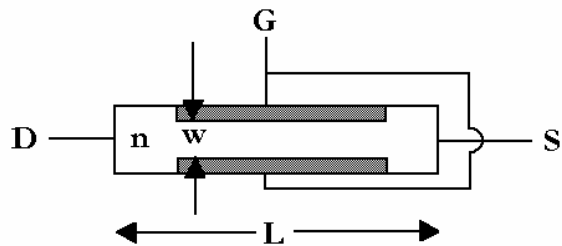

Figure 1. Cross-sectional view of a typical field effect transistor.

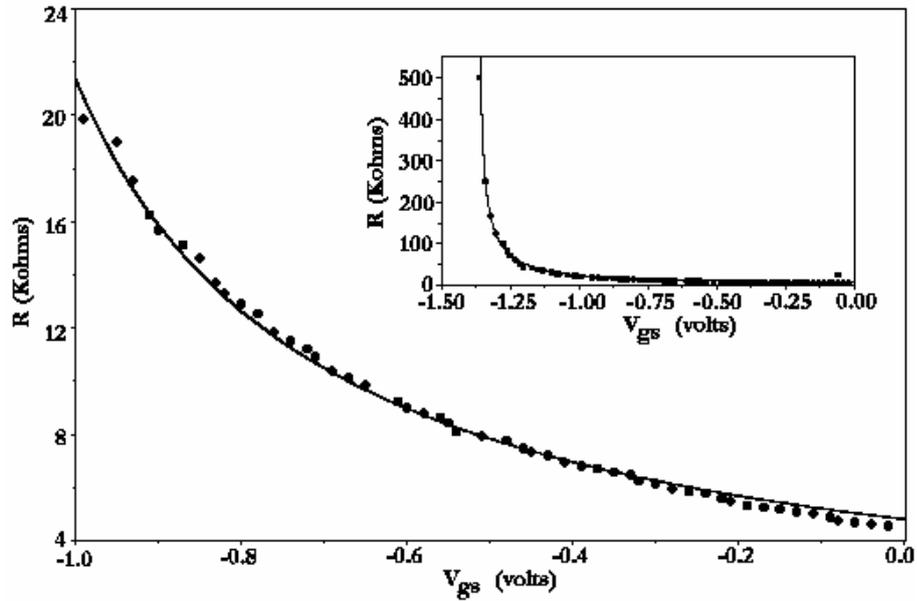

Figure 2. The channel resistance as measured with varying gate-source voltage. The inset gives the complete variation from 0 to -1.5 volts.

The channel resistance is a function of the gate-source voltage.[1] Increased reverse biasing increases the resistance. Figure 2 shows the measured variation in channel resistance with gate-source voltage ($V_{gs}$). Figure 3 shows the circuit used to experimentally determine the FET's resistance as a function of it gate-source voltage. The resistance is calculated from the experimentally determined $V_o$, using the expression

$$R_{fet} = \left(\frac{5-V_0}{V_0}\right) 1\text{K}\Omega$$

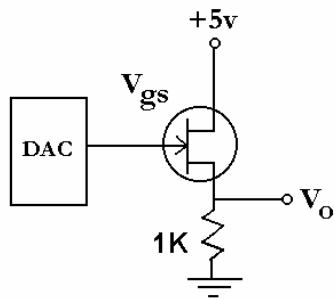

Figure 3. The experimental set-up used to measure the FET's resistance as a function of gate-source voltage.

This feature of the FET gives rise to an unique application for the device. We have used a computer to send various digital outputs to a digital to analogue converter (DAC) whose output in turn is given as $V_{gs}$ to the FET. The FET was used as a resistance in one of the arms of an inverting op-amp amplifier[2] (see Figure 4). The gain of the circuit is given as

$$A_v = -\frac{R_f}{R_1}$$

Where symbols have standard meanings. The FET can be used as a resistance either in place of $R_1$ or $R_f$. The varying FET resistance resulted in varying gain of the amplifier. Thus, we were able to design a *programmable amplifier*.

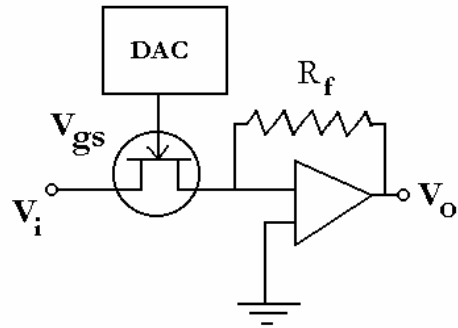

Figure 4. The inverting amplifier circuit where resistance $R_1$ has been replaced by an FET as a programmable resistance.

The experiment allows the students to view the FET in a new perspective and adds a new experiment which exposes students to a possible application of the device. Inquisitive students, with a little help, can design programmable circuits combining their knowledge of computers, digital electronics and analog electronics.

References

1.  P.Arun, *Electronics*, (Narosa) New Delhi 2005.
2.  R. Gayakwad, *Op amps and Linear Integrated Circuits*, (PHI) New Delhi 1989.